\magnification=1200
\def\setup{\count90=0 \count80=0 \count91=0 \count85=0
\countdef\refno=80 \countdef\secno=85 \countdef\equno=90
\countdef\ceistno=91 }
\def\R{\vrule height5.85pt depth.2pt \kern-.05pt \tt R}

\def\Box{\vbox{\hrule
                    \hbox{\vrule height 6pt \kern 6pt \vrule height 6pt}
                    \hrule}\kern 2pt}
\def\lo{\raise2pt\hbox{$<$}\kern-7pt\raise-2pt\hbox{$\sim$}}
\def\go{\raise2pt\hbox{$>$}\kern-7pt\raise-2pt\hbox{$\sim$}}
\def\parallel#1#2{\hbox{\kern1pt \vrule height#1pt
                                 \kern#2pt
                                 \vrule height#1pt \kern1pt}}

\def\sub#1{{\lower 8pt \hbox{$#1$}}}

\def\circum#1{{ \kern -3.5pt $\hat{\hbox{#1}}$ \kern -3.5pt}}
\def\Del{{\raise.5ex\hbox{$\bigtriangledown$}}}
\def\DEL#1{{\raise.5ex\hbox{$\bigtriangledown$}\raise 8pt \hbox{\kern -10pt
                 \hbox{$#1$}} }}
\def\autoeq{ {\global\advance\count90 by 1} \eqno(\the\count90) }
\def\autoeql{ {\global\advance\count90 by 1} & (\the\count90) }
\def\ceist{ {\global\advance\count91 by 1} (\the\count91) }
\def\autosec{ {\global\advance\secno by 1} (\the\secno) }
\def\e{\hbox{e}}
\def\Lie#1{{\cal L}{\kern -6pt 
            \hbox{\raise 1pt\hbox{-}}\kern 1pt} _{\vec{#1}}}

\def\Z{Z \kern-5pt \hbox{\raise 1pt\hbox{-}}\kern 1pt}
\def\eye{\bigcirc\kern -7pt\bigcirc}

% THE FOLLOWING MACROS ARE FOR AUTOMATICALLY ARRANGING REFERENCES
% TO GET A REFERENCE USE \refread{#}
% IT REQUIRES A SINGLE ARGUMENT, WHICH IS THE REFERENCE # IN refs.tex
% ONCE \readref HAS BEEN INVOKED THE TITLE RESIDES IN \title
% THE AUTHOR IN \author AND THE DETAILS OF YEAR, JOURNAL OR
% PUBLISHER, VOLUME AND PAGE # IN \pub

\def\autoref{ {\global\advance\refno by 1} \kern -5pt [\the\refno]\kern 2pt}

\def\readref#1{{\count100=0 \openin1=refs\loop\ifnum\count100
<#1 \advance\count100 by1 \global\read1 to \title \global\read1 to \author
\global\read1 to \pub \repeat\closein1  }}

\def\reftitle{{ \kern -3pt \vtop{ \hbox{\title} \hbox{\author\ \pub} } }}
\def\ref{{ \kern -3pt\author\ \pub \kern -3.5pt }}

\def\refanon{{ \hbox{\pub}\kern -3.5pt }}

\def\reflongtitle{{ \kern -3pt \vtop{ \hbox{\title} 
                                      \hbox{\author\hfill}
                                      \hbox{\pub}    }   }}
\input epsf
\overfullrule=0pt
\setup
\hsize 130mm
\parindent0pt
\centerline{  }
%\line{\hbox{DIAS-STP-94-39 }\hfill \hbox{Fifth Revision}}
%\line{\hbox{cond-mat/9412031}\hfill \hbox{1st February 1996 }}
\vskip 2cm  
\centerline{\bf CHAOTIC BEHAVIOUR OF RENORMALISATION FLOW IN} 
\centerline{\bf A COMPLEX MAGNETIC FIELD}
\vskip 1.2cm
\centerline{Brian P. Dolan}
\vskip .5cm
\centerline{\it Department of Mathematical Physics, St. Patrick's College}
\centerline{\it Maynooth, Ireland}
\centerline{and}
\centerline{\it Dublin Institute for Advanced Studies}
\centerline{\it 10, Burlington Rd., Dublin, Ireland}
\vskip .5cm
\centerline{e-mail: bdolan@thphys.may.ie}
\vskip 1.5cm
\centerline{ABSTRACT}
\noindent It is demonstrated that decimation of
the one dimensional Ising model, with periodic boundary conditions,
results in a non-linear renormalisation transformation for the couplings
which can lead to chaotic behaviour when the couplings are complex. The
recursion relation for the couplings under decimation is equivalent
to the logistic map, or more generally the Mandelbrot map.
In particular an imaginary external magnetic field can give chaotic 
trajectories in the space of couplings.
The magnitude of the field must be greater than a minimum value
which tends to zero as the critical point $T=0$ is approached,
leading to a gap equation and associated critical exponent which are 
identical to
those of the Lee-Yang edge singularity in one dimension. 
\vskip .5cm \noindent
% PACS Nos. $0530$, $0545$  and $0550$
\vfill\eject
%\baselineskip = 1.5\baselineskip
The renormalisation group has been developed into an immensely powerful tool
for the analysis of physical theories near critical points and also for continuum
field theories.  There are by now various forms of \lq\lq renormalisation group
equation" 
which govern how physical amplitudes and couplings change under
change of scale. Perhaps one of the most intuitively appealing is the version
due to Wilson
\autoref\newcount\Wilson\Wilson=\refno,
motivated by a suggestion of Kadanov
\autoref\newcount\Kadanov\Kadanov=\refno,
involving 
\lq\lq decimation".
\bigskip
In principle one can derive recursive formulae for the couplings of a theory, which dictate how
they should change when the underlying lattice is decimated, so that the Hamiltonian
involving the new couplings on the new lattice is the same as the Hamiltonian
involving the old couplings on the old lattice, i.e. the partition function 
does not change under the simultaneous operations of decimation and 
redefinition of couplings.
The recursive formulae for the couplings are in general non-linear (indeed 
they are not invertible, so the transformation involved here is not a group
but a semi-group).  
\bigskip  
Non-linear recursive formulae are one of the central themes of 
study for chaos theory and one can pose the question - can the renormalisation
transformation lead to chaotic behaviour in the space of couplings?  
This possibility has been investigated before, and answered in the
affirmative using numerical calculations in some specific models
for which exact recursion relations can be obtained, 
\autoref\newcount\McKay\McKay=\refno  
\autoref\newcount\Svrakic\Svrakic=\refno  
\autoref\newcount\Derrida\Derrida=\refno
\autoref\newcount\Damgaard\Damgaard=\refno.  
Here a simple model (the one dimensional Ising model)
will be analysed analytically
and it will be shown that
this model
also exhibits chaotic behaviour in a surprisingly elegant manner.
The analysis shows that the onset of 
chaotic behaviour appears to be associated with the second order phase transition
at T = 0.  It remains an open question as to whether this is a peculiarity of
this model or is a more general feature.
\bigskip   
One severe problem in extracting general features is the paucity of models
for which the recursion relations can be obtained exactly, and the one dimensional
Ising model - because of its simplicity - is one example for which progress
can be made.  Nevertheless, despite its simplicity, the results are startling
enough to merit description.
\bigskip  
It will be shown that the onset of chaotic behaviour is brought about by extending
the couplings of the theory to the complex plain.  This is not a new idea in
the analysis of such theories.  Dyson
\autoref\newcount\Dyson\Dyson=\refno
pointed out that one could learn
something about the structure of Quantum Electrodynamics by considering 
imaginary electric charges, so that $\alpha=e^2/ \hbar c < 0$.  Such a theory
must be intrinsically unstable, and so amplitudes cannot be analytic at 
$\alpha=0$, hence perturbation theory must diverge and expansions in $\alpha$ are,
at best, asymptotic.
These ideas have been further developed by making $\alpha$ complex and there
is by now a whole literature on complex analyticity and Borel summability
(e.g.\autoref\newcount\Sokal\Sokal=\refno).  
In statistical mechanics, extending the couplings to the complex
plane is a key step in solving many two-dimensional 
models\autoref\newcount\Baxter\Baxter=\refno
and has led
to some beautiful results concerning the analyticity of the partition function
\autoref\newcount\LeeYang\LeeYang=\refno.
\bigskip  
In this paper yet another example of the fascination of complex variables 
will be exhibited - by
allowing the couplings of the one dimensional Ising model to be complex, the
recursive renormalisation transformations can become chaotic.  To exhibit
this phenomenon, some well known features of the one dimensional Ising model
will be summarised and the recursion relations derived.  It will then be shown
that the recursion relation is nothing other than the logistic map, and chaos
ensues!
\bigskip  
Consider the one dimensional Ising model on a periodic lattice of N sites
[\the\Baxter].  The partition function is
$$
Z_N=\sum_{\{\sigma\}} exp \left[ {K \sum_{j=1}^N \sigma_j \sigma_{j+1} + h 
\sum_{j=1}^N \sigma_j} \right]  
\autoeq
$$

where $K = {{J} \over {kT}}$ and $h = {{H} \over {kT}}$, with J the spin coupling 
and H the external magnetic field, (periodic boundary conditions require 
$
\sigma_{N+1} \equiv \sigma_1)
$. 
$Z_N(K,h)$
can be conveniently expressed in terms of the transfer matrix
$$
V=\pmatrix{
V_{++}&V_{+-}\cr
V_{-+}&V_{--}\cr}
\quad = \quad
\pmatrix{
e^{K+h}&e^{-K}\cr
e^{-K}&e^{K-h}\cr
} \autoeq
$$

as $Z_N = TrV^N$.
\bigskip  
Diagonalising $V$ gives the eigenvalues
$$
\lambda_\pm=e^K\left\{\cosh h\pm\sqrt{\sinh^2 h+e^{-4K}}\right\}.
\autoeq
$$ \newcount\Eigenvalues\Eigenvalues=\count90
Thus 
$$
Z_N=\lambda_+^N\left[{1+}\biggl({\lambda_- \over \lambda_+}\biggr)^N\right].
\autoeq
$$
\bigskip  
The recursive renormalisation transformation is well known for
the 1-D Ising model
\autoref\newcount\Fisher\Fisher=\refno. 
It is obtained by asking: can
one find new couplings $K^\prime$ and $h^\prime$ such that
$$
Z_{N\over2}(K^\prime,h^\prime)=A^NZ_N(K,h)
\autoeq 
$$\newcount\part\part=\count90  
gives the same physical amplitudes? ($A$ is a normalisation factor.)                                    
\bigskip
Equation (\the\part) is easily satisfied by demanding
$$
\pmatrix{
e^{K^\prime+h^\prime}&e^{-K^\prime}\cr
e^{-K^\prime}&e^{K^\prime-h^\prime}\cr
}
\quad = A^2\quad
\pmatrix{
e^{K+h}&e^{-K}\cr
e^{-K}&e^{K-h}\cr
}^2 \autoeq
$$
giving the recursive formulae:
$$
{e^{2h^\prime}=e^{2h}}\quad {\cosh (2K+h)\over\cosh(2K-h)}
\autoeq
$$\newcount\newconst\newconst=\count90

$$
e^{4K^\prime}=\quad {\cosh(4K)+\cosh(2h)\over 2\cosh^2(h)}.
\autoeq$$\newcount\recrel\recrel=\count90
The normalisation factor $A$ is unimportant for the present analysis.
\bigskip
The combination $e^{4K^\prime}\sinh^2(h^\prime) = e^{4K} \sinh^2(h)$ is a renormalisation
transformation invariant.  This means that the magnetisation per
site, in the thermodynamic limit, 
$$
M_{N\rightarrow\infty}={\partial \, ln \, Z_N \over 
N \partial h}= {e^{2K}\sinh(h)
\over\sqrt{1+e^{4K}\sinh^{2}(h)}} 
\autoeq
$$\newcount\Mag\Mag=\count90
is invariant under the renormalisation transformation.  
\bigskip
All these facts about the one dimensional Ising model are well known 
[\the\Baxter]
and are included only for completeness.
\bigskip
It will now be shown that the recursion relations (\the\newconst) and
(\the\recrel) are equivalent to the
logistic map and, for certain (complex) values of couplings gives rise to chaotic
behaviour.
\bigskip
Define
$$
m=1+e^{4K}\sinh^2(h)
\autoeq
$$\newcount\mdef\mdef=\count90
which is a renormalisation transformation invariant, $m=m^\prime$.  It is now only
necessary to consider one of equations (\the\recrel) and 
(\the\newconst) as the existence of the
invariant, $m$, makes one of them redundant.  
\bigskip
Eliminating $h$ from (\the\recrel) using (\the\mdef) gives
$$
e^{4K^\prime} -1 = {1\over4} {(e^{4K}-1)^2 \over [(e^{4K}-1)+m]}.
\autoeq 
$$\newcount\singrec\singrec=\count90
Now replace $K$ with a new variable
$$
x  = - {m \over {(e^{4K}-1)}}
\autoeq
$$
with -$\infty < x < 0$  for $m > 0$  and $K > 0$.
The recursion relation (\the\singrec) now becomes
$$
x^\prime=4x(1-x)
\autoeq
$$\newcount\logmap\logmap=\count90
which is the logistic map.
\bigskip
For $0 < x < 1$, the recursion relation (\the\logmap) leads to chaotic behaviour, 
as is easily seen by defining $x=\sin^2(\pi\psi)$, $0<\psi<{1\over2}$ 
(see e.g.\autoref\newcount\Schroeder\Schroeder=\refno)
giving 
$$
\sin(\pi\psi^\prime)=\sin(2\pi\psi)
\autoeq
$$
Writing $\psi$ in binary form, we see that the iterative map merely shifts
all bits one step to the left and throws away the integral part, leaving the 
fractional part behind.  For an initial value of $\psi$ which is rational this
will lead to a periodic orbit, but for a starting value of $\psi$ which is 
irrational, the process never repeats and $\psi$ jumps around chaotically.
Since the irrationals have a greater cardinality than the rationals,
almost all initial values lead 
to chaotic motion.
\bigskip
For real values of the couplings, $m > 1$ and -$\infty < x < 0$.  Chaotic
trajectories require 
$$
m=1+e^{4K}\sinh^2(h)<0.
\autoeq
$$
For example, if $K>0$ is real and $h$ is pure imaginary $(h=i\theta)$, then
$m<0$ for $\sin^2\theta>e^{-4K}$
and
$x={e^{4K}\sin^2\theta-1 \over e^{4K}-1}$ lies between 0 and 1
for
$e^{4K}>{1\over{\sin^2\theta}}$.
\bigskip
The region of chaotic flow is shown in Figure 1. Note that not all
points above the line $x=0$ lead to chaotic trajectories, only those with
irrational $\psi$. Indeed there is an infinite number of periodic
trajectories as well as chaotic ones. 
For example there are lines of unstable fixed points (period one) at 
$x=0$ and $x=3/4$ and
the values $x=(5\pm\sqrt{10})/8$
give orbits of period two , etc.
\bigskip
For finite K there is a gap and a small imaginary magnetic field is not
sufficient to induce chaos, but as K increases $(T\rightarrow 0)$ this gap
reduces to zero.
Following Baxter [\the\Baxter], 
define $t=e^{-2K}$ with $t\rightarrow 0$ being the critical
point, then near $t=0$ the line separating chaotic from regular flow is given by
$$
t^2\sim\theta^2\quad \Rightarrow \quad \theta\sim t.
\autoeq
$$
If we define a critical exponent, $\Delta$, such that the critical value of 
$\theta$ is 
$$
\theta\sim t^\Delta,
\autoeq
$$
then one obtains $\Delta = 1$ for the one dimensional Ising model,
which is just the critical exponent for the Lee-Yang edge singularity
in this model.
It is not difficult to see that the critical line 
$e^{4K}\sin^2\theta=1$ is related to the Lee-Yang zeros of the partition
function in the complex $h$-plane. The equation $Z_N(K,h)=0$ has
$N$ roots, for real $K$,
$$Z_N = (\lambda_+)^N +(\lambda_-)^N = 0 \qquad\Leftrightarrow \qquad 
\lambda_+ = \e^{iq\pi\over N}
\lambda_-,\autoeq$$
where $-N<q\le N$ is odd.
Using the explicit form  of the eigenvalues, (\the\Eigenvalues), this leads to
$$\cos\Bigl({q\pi\over 2N}\Bigr)\sqrt{\e^{-4K} + \sinh^2(h)} 
=i\sin\Bigl({q\pi\over 2N}\Bigr)\cosh(h).\autoeq$$
This equation can be re-arranged to give
$$\cos(\theta_q)=\sqrt{1-t^2}\cos\left({q\pi\over 
2N}\right),\autoeq$$
where $t=\e^{-2K}$ as before and $\theta_q$ are the $N$ roots in the
rotated complex $h$-plane, $h=i\theta$.
Since $0<t<1$ the $N$ values of $\theta_q$ are all real
and they lie in the range $t<\vert\sin(\theta_q)\vert<1$,
which is precisely the region above the critical line in figure 1.
In the thermodynamic limit ($N\rightarrow\infty$) 
the lowest zeros are at $\sin(\theta_q)=t$ which is 
the critical line. Thus the critical line co-incides with the
rightmost Lee-Yang zeros in the complex activity plane in the 
thermodynamic limit. 

One can obtain further insights by allowing K to become complex.
Define $x=-{z\over4}+{1\over2}$ in equation (\the\logmap) to give 
$$
z^\prime=z^2-2.
\autoeq
$$
This is the Mandelbrot map $z^\prime=z^2+c$ for complex $z$, 
with $c=-2$.
One can have 
divergence or convergence depending on $c$ and the initial choice of $z$.
The values of $c$ for which the iterates of the starting point $z=0$ 
stay bounded is the Mandelbrot set and clearly
$c=-2$ is an element of this set. The Julia
set for a given value of $c$ is the set of points in the complex $z$-plane
which stay within a bounded region upon repeated iteration of the
Mandelbrot map (strictly speaking this is the filled in Julia set $J_c$ - the
Julia set $J$ is actually the boundary of this set). The set $J$ is
generated by the inverse set of the unstable fixed points. For $c=-2$ these
are $z=-1$ and $z=2$ and so the
earlier analysis of the logistic map tells us that the
inverse iterates generate a dense set of points in the segment
of the real axis lying between $-2$ and $+2$
(this corresponds to $0<x<1$ in the previous notation). Thus 
the filled in Julia set is just the segment of the real axis with
$-2< z <2$.
This analysis shows that 
the forward iterations send $\vert z\vert$ to infinity if the
temperature has an imaginary component or if the magnetic field
has both real and imaginary parts non-zero (this latter
possibility would result in $m$ having non-zero imaginary
part and thus so would $z$). 
Thus chaotic trajectories occur only
for real $K$ and pure imaginary $h$. The behaviour of the Julia set
under iteration is shown in Figure 2 for $c=-2$.
\bigskip
An obvious question is: how generic is this behaviour?  For a
general Hamiltonian when is it 
possible to obtain chaotic behaviour in some region of (complex) 
coupling space?  For the moment this question must remain unanswered, but
a few comments should be made.  Feigenbaum was aware of the universality
in chaos
\autoref\newcount\Feigenbaum\Feigenbaum=\refno.  
Near an extremum 
any non-linear map (with non-vanishing second derivative)
can be put into the form (\the\logmap) 
with the number
4 replaced, in general, by a parameter, $\lambda$.  
\bigskip
Thus
$$
x^\prime=\lambda x(1-x)
\autoeq
$$
is generic, but whether or not one has chaotic behaviour depends on the
value of $\lambda$ and the initial value of $x$. 
I do not know of any reason why $\lambda$ has the  
rather special value of 4 for the one dimensional Ising model.  
More generally, Feigenbaum has shown [\the\Feigenbaum] that the
properties of a general non-linear map
$$x^\prime = \lambda f(x)\autoeq$$
are independent of the exact form of $f(x)$ near a maximum.
For any particular model, the value of $\lambda$ would have to be calculated 
ab initio and I know of no way of deciding in advance whether or not chaotic 
flow would result. For the 1-D Ising model it is clear that the
onset of chaotic trajectories is related to the second order phase
transition at $T=0$ and the existence of the Lee-Yang edge singularity.
However other models show chaotic recursive maps for real
values of the couplings which are not related to second order phase
transitions - rather they are related to frustration and glass-like
structures [3]. Thus it does not appear that a second order phase
transition is a pre-requisite for chaos, but neither are spin-glasses
a pre-requisite.
Unfortunately the number of models for which the 
recursion relations are known exactly is rather few. In particular
the recursion relations for the 2-D Ising model are not known, so
it is not possible at this stage to say whether or not the two
dimensional Lee-Yang edge singularity is related to chaotic trajectories.
\bigskip
It is a pleasure to thank Chris Stephens and Bill McGlinn for
useful discussions on the one dimensional Ising model,
Tony O'Farrell for helpful discussions about the Julia set
and Paul Upton and Denjoe O'Connor for invaluable suggestions
concerning Lee-Yang zeros.
\bigskip

\vfill\eject
{\bf References}\hfil
\vskip .5cm
\item{[\the\Wilson]} K.G. Wilson and J. Kogut, Phys. Rep. C {\bf 2} (1974) 75
\smallskip
\item{[\the\Kadanov]} L. Kadanoff, Physica {\bf 2} (1966) 263
\smallskip
\item{[\the\McKay]} S. MacKay, A.N.~Berker and S.~Kirkpatrick, Phys. Rev. Lett.
{\bf 48}, 767, (1982); A.N.~Berker and S.~McKay, J. Stat. Phys.
{\bf 36}, 787, (1984)
\smallskip
\item{[\the\Svrakic]} N.M. Svrakic, J. Kertesz and W. Selke, J. Phys. A {\bf 15},
L427, (1982)
\smallskip
\item{[\the\Derrida]} B. Derrida, J.-P. Eckmann and A. Erzan. J. Phys. A {\bf 16},
893, (1983)     %Logistic Map, Analyticity of free energy for chaotic systems
\smallskip
\item{[\the\Damgaard]} P.H. Damgaard and G. Thorleifsson, Phys. Rev. A {\bf 44}, 2738,
(1991);\hfill\break
P.H. Damgaard, Int. J. of Mod. Phys. A {\bf 7}, 6933, (1992)
\smallskip
\item{[\the\Dyson]} F.J. Dyson, Phys. Rev. {\bf 85} (1952) 631
\smallskip
\item{[\the\Sokal]} A.D. Sokal, J. Math. Phys. {\bf 21} (1979) 261
\smallskip
\item{[\the\Baxter]} R.J. Baxter, {\sl Exactly Solved Models In 
Stastitical Mechanics}\hfill\break
 Academic Press (1982)
\smallskip
\item{[\the\LeeYang]} C.D. Yang and T.D. Lee, Phys. Rev. {\bf 87} (1952) 
404,\hfill\break 
Phys. Rev. {\bf 87} (1952) 410
\smallskip
\item{[\the\Fisher]} M.E. Fisher, {\sl Scaling Universality and
Renormalisation Group Theory  \hfill\break}
Proc. of the Univ. of Stellenbosch Summer School, South Africa,
1982\hfill\break 
Springer Lecture Notes in Physics {\bf 186} (1983)
\smallskip
\item{[\the\Schroeder]} M. Schroeder, {\sl Fractals, Chaos, Power Laws}
W.H. Freeman and Co. (1991)
\smallskip
\item{[\the\Feigenbaum]} M.J Feigenbaum, J. Stat. Phys. {\bf 21} (1979) 669
\vfill\eject
\centerline{\bf Figure Captions}
\bigskip
\item{Figure 1.} Critical line in the $K$-$\theta$ plane. The renormalisation
flow is regular below the critical line $x=0$ and chaotic above it for
all values of $\psi$ which are irrational ($x=\sin^2(\pi\psi)$).
\bigskip
\item{Figure 2.} A representation of the Julia set in the complex $z$-plane
for $c=-2$. The
Julia set itself is the real line segment $-2<\hbox{Re}(z)<2$ which is
the width of the diagram. The contours
depict the rate at which a point is repelled from the Julia set, the 
darker the
contour the less rapid the expulsion. The picture was generated use
the program FRACTINT produced by the Stone Soup Group.
\vfill\eject
\nopagenumbers
\epsfxsize=115mm
\epsffile{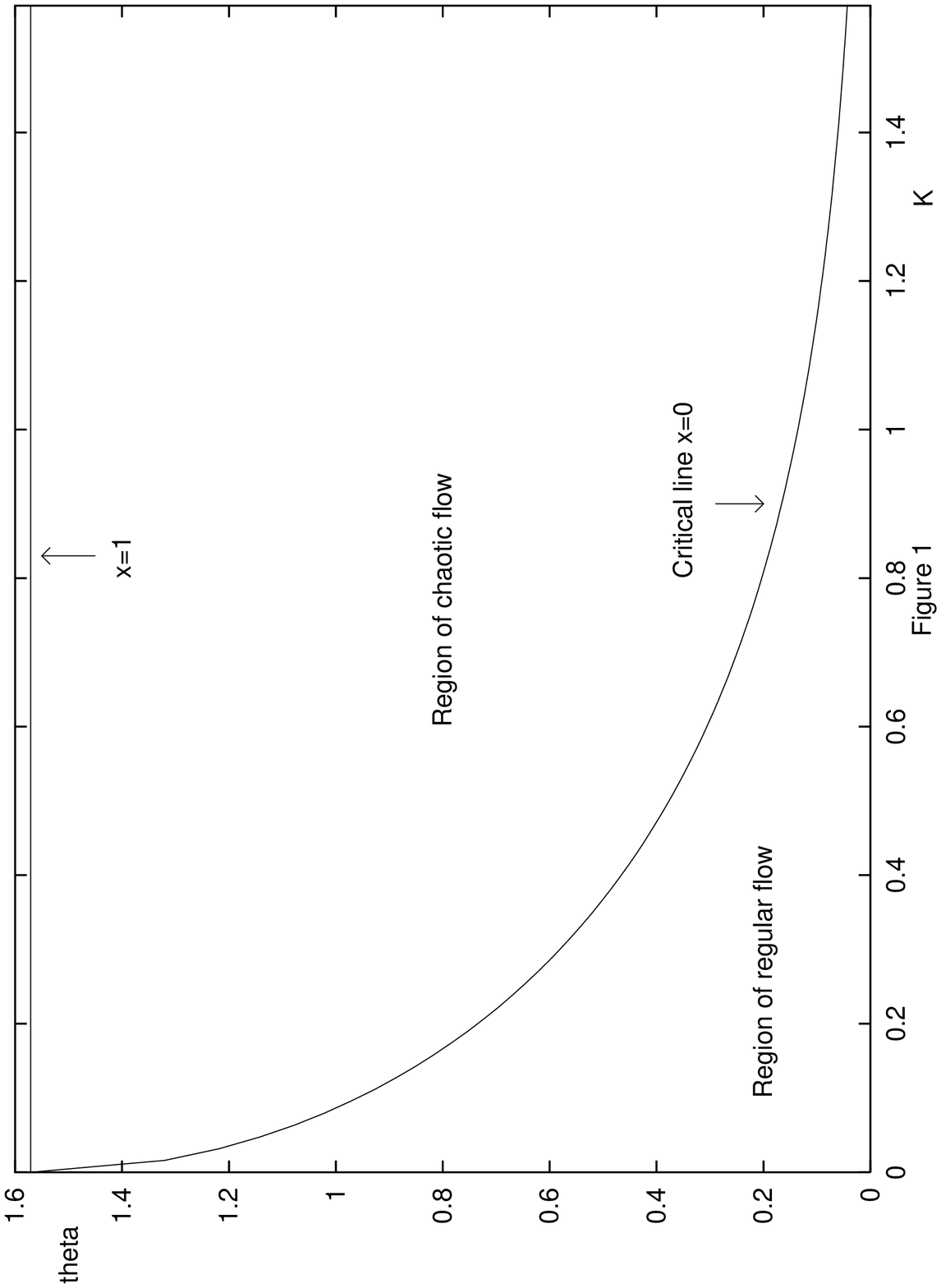}
\vfill\eject
\epsfxsize=115mm
\epsffile{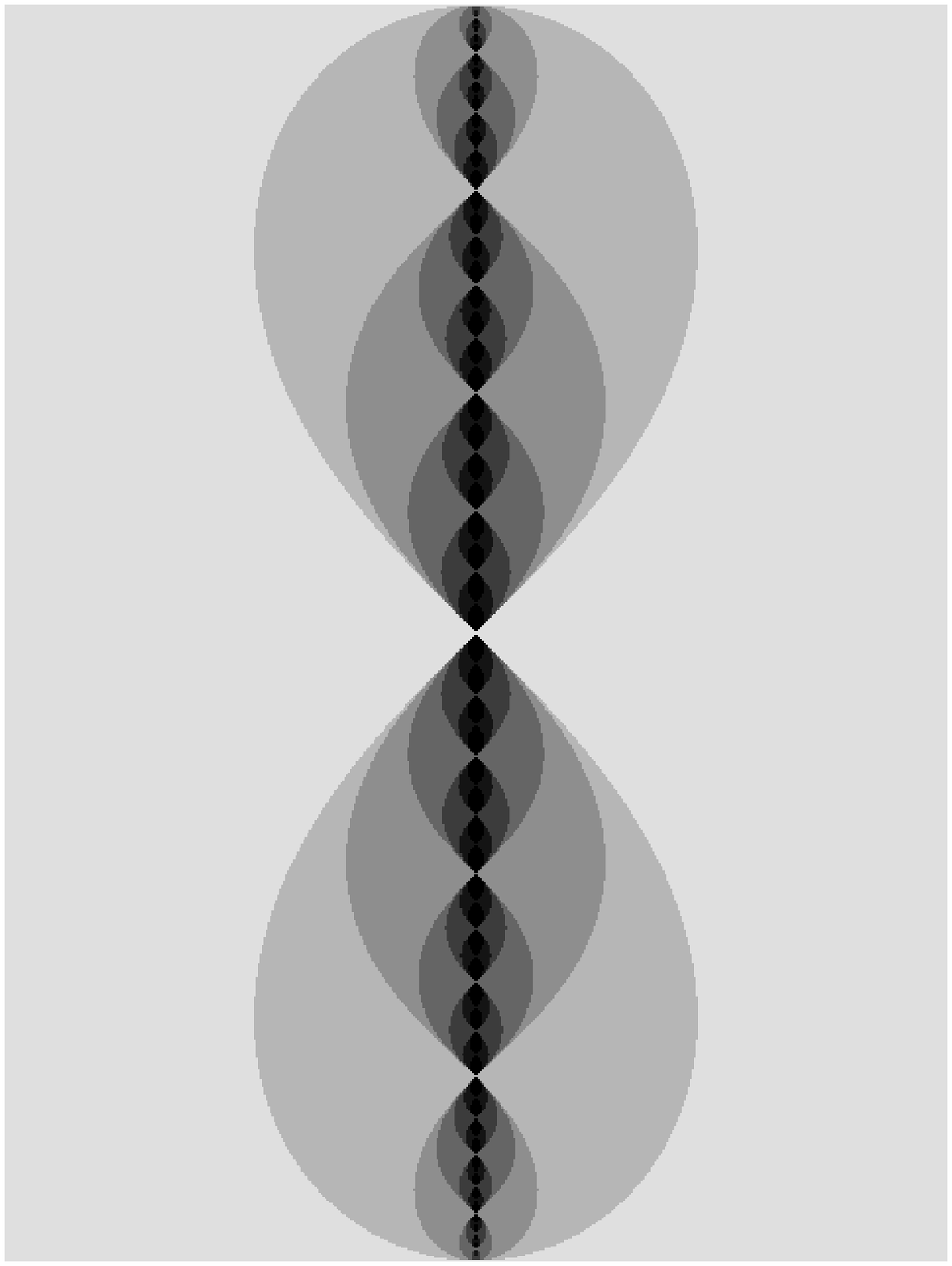}
\bye